\documentclass[aip,sd,
amsmath,amssymb,
 reprint,%
]{revtex4}

\usepackage{graphicx}% Include figure files
\usepackage{dcolumn}% Align table columns on decimal point
\usepackage{bm}% bold math
\newcommand{\etal} { et~al.\ }  
\newcommand{\sol}{$M_\odot$}
\newcommand{\araa}{Ann. Rev. Astron. Astrophys.}
\newcommand{\apss}{Astrophys. \& Space Sci.}
\newcommand{\apjl}{Astrophys. J. Letters}
\newcommand{\aanda}{Astronomy \& Astrophys.}
\newcommand{\physflu}{Physics of Fluids}
\newcommand{\jfm}{J. Fluid Mech.}
\newcommand{\mnras}{Mon. Notices R.A.S.}
\newcommand{\zap}{Zeits. f\"ur Astrophysik}

\def\favg#1{\widetilde{#1}} 
\def\ff#1{#1''} 
\def\favgDt{\favg{D}_t} 
\def\eavg#1{\overline{#1}} 
\def\erho{\eavg{\rho}}

\begin{document}

%\preprint{AIP/123-QED}

\title{Chaos and turbulent nucleosynthesis prior to a supernova explosion}

\author{W. D. Arnett}
% \altaffiliation[Also at ]{Physics Department, XYZ University.}
 \email{darnett@as.arizona.edu.} 
\author{C. Meakin}%
 \email{cmeakin@as.arizona.edu.}
\affiliation{ Steward Observatory, University of Arizona, 
933 N. Cherry Avenue, Tucson AZ 85721
%\\This line break forced with \textbackslash\textbackslash
}%

\author{M. Viallet}
% \homepage{http://www.Second.institution.edu/~Charlie.Author.}
\email{mviallet@mpa-garching.mpg.de}
\affiliation{%
Max-Planck Institut f\"ur Astrophysik, Karl Schwarzschild Strasse 1, Garching, D-85741, Germany 
%\\This line break forced% with \\
}%

\date{\today}% It is always \today, today,
             %  but any date may be explicitly specified

\begin{abstract}
Three-dimensional (3D), time dependent numerical simulations of flow of matter in stars, now have sufficient resolution to be fully turbulent. The late stages of the evolution of massive stars, leading up to core collapse to a neutron star (or black hole), and often to supernova explosion and nucleosynthesis, are strongly convective because of vigorous neutrino cooling and nuclear heating. Unlike models based on current stellar evolutionary practice, these simulations show a chaotic dynamics characteristic of highly turbulent flow.
Theoretical analysis of this flow, both in the Reynolds-averaged Navier-Stokes (RANS) framework and by simple dynamic models, show an encouraging consistency with the numerical results.  It may now be possible to develop physically realistic and robust procedures for convection and mixing which (unlike 3D numerical simulation) may be applied throughout the long life times of stars.  In addition, a new picture of the presupernova stages is emerging which is more dynamic and interesting (i.e., predictive of new and newly observed phenomena) than our previous one.

%Valid PACS numbers may be entered using the \verb+\pacs{47.27}+ command.
\end{abstract}

\pacs{Valid PACS appear here}% PACS, the Physics and Astronomy
                             % Classification Scheme.
\keywords{Suggested keywords}%Use showkeys class option if keyword
                              %display desired
\maketitle

%\begin{quotation}
%The ``lead paragraph'' is encapsulated with the \LaTeX\ 
%\verb+quotation+ environment and is formatted as a single paragraph before the first section heading. 
%(The \verb+quotation+ environment reverts to its usual meaning after the first sectioning command.) 
%Note that numbered references are allowed in the lead paragraph.
%%
%The lead paragraph will only be found in an article being prepared for the journal \textit{Chaos}.
%\end{quotation}

\section{\label{sec:level1}Introduction}
With the continuing increase of computing power, it has become possible to numerically simulate --- in three dimensions (3D) and with sufficient resolution to allow turbulent flow --- aspects of the last stages of evolution of a massive star before core collapse. All such simulations show dynamic behavior beyond that contained in 1D stellar evolutionary models. Some of the dynamic behavior is intrinsic to turbulent convection, and some may be due to interactions between active burning shells. Because the star arranges itself to maintain an approximate balance between cooling by neutrino emission and heating by thermonuclear burning, the stages of burning of carbon, neon, oxygen and silicon are made shorter and more vigorous than they would otherwise be.
These stages are crucial for nucleosynthesis,  they set the initial conditions for core collapse to form a neutron star or black hole, and they affect any ensuing supernova explosion.  It now appears that current ideas about these stages are quantitatively, and perhaps qualitatively, incorrect due to a flaw in the physics used in the 1D stellar evolutionary models: the treatment of convection and mixing. 

\subsection{A History of Convection in Stellar Models}

The detailed numerical modeling of stars begins with work on mechanical calculators by Schwarzschild (1957) and Hoyle (1955), and flourished with the application of digital computers in the work of Henyey, et al. (1959), Iben (1965, 2013), Kippenhahn \& Wiegert (1990) and others. These efforts contained the assumption that mixing times were short in comparison to nuclear burning times, so that convection zones were completely mixed (homogeneous in composition).
A key problem was the choice of method used to join the nearly adiabatic convection of the deep interior with the radiatively dominated, nonadiabatic convection which occurred as the stellar surface was approached.
Erika B\"ohm-Vitense (1958, also Vitense 1953) provided an algorithm based on the mixing length ideas of Prandtl, which became the standard in the stellar evolution community, and known as ``mixing-length theory" (MLT). MLT has one parameter which is actually adjusted, and in this sense is a one parameter theory; this adjustment is done so that a solar model would have the observed radius of the present day sun. Adjustment to get the observed luminosity constrained the solar abundances, at least to the extent that MLT was a physically correct description of solar convection. B\"ohm-Vitense invented MLT prior to two major discoveries relevant to turbulent flow:  the work on the turbulent cascade appeared in the western literature in Kolmogorov (1962), and work on chaos in a convective roll in Lorenz (1963). A great success of MLT was the description of the ``ascent of the giant branch", in which stars evolve to larger radii and luminosity, with vigorous surface convection regions, using essentially the same value of the mixing length parameter as found from solar calibration. 

What about more advanced stages of stellar evolution?
Rakavy, Shaviv and Zinamon (1967) examined the evolution of C and O cores in neutrino cooled stages up to the onset of core collapse instability, assuming complete mixing inside convective zones.
Arnett (1972) considered He cores (which formed C and O cores by burning He, giving a connection to photon cooled stages), introduced time dependent convection and mixing as an advective process, and evolved the cores into collapse and neutrino trapping (Arnett 1977).
For shell burning, Eggleton (1972) introduced the notion of ``convective diffusion" in which the advective action of turbulent convection is approximated by a diffusion operator, and emphasized that this was merely a numerical convenience (see $\S $\ref{sec:level1}). 
Weaver, Zimmerman, \& Woosley (1978) extended this notion to late burning stages of massive stars. 
Langer, El Eid \& Fricke (1985) considered semiconvection as well, and offered a modified diffusion coefficient.
 This became popular and is now regarded as the standard in the field (Langer, 2012), although we could find no convincing justification for approximating a turbulent process by an heuristic diffusion in the astronomical literature. We note that
 Daly \& Harlow (1970) derive the conditions for such diffusion, which include the constraint that the convective scale be small relative to the integral scale;  this is not true in the stellar case, for which they are comparable.

%langer araa, maedre-meynet, limongi-chieffi, heger-woosley, Hirschi 

 \subsection{Computer Simulations of Turbulence in Multidimensions}	
Flows in stellar plasma are highly turbulent. The parameter which indicates the onset of turbulence is the
Reynolds number 
\begin{equation}
{\cal R}_e= \ell  u/\nu,
\end{equation}
where $\ell$ is a length characterizing the flow, $u$ a velocity scale characterizing the flow, and $\nu$ the kinematic viscosity. The Reynolds number is the ratio of inertial forces to viscous forces. As a representative example we use the Spitzer-Braginskii  estimate of kinetic viscosity of a classical plasma (Braginskii 1958; Spitzer 1962, p.~146),
\begin{equation}
\nu = 2.21 \times 10^{-15} \ T^{5/2} A^{1/2} /( \rho Z^4 \ln \Lambda) \ \rm cm^2/sec.
 \end{equation}
 The coulomb logarithm is often taken to be $\ln \Lambda \sim 10$.
 %\begin{equation}
%\nu = 6.99 \times 10^{2} \ T_7^{5/2} A^{1/2} /( \rho Z^4 \ln \Lambda) \rm cm^2/sec
 %\end{equation}
For stellar conditions, $\nu \sim 70$, to be compared with
$\nu \sim 1/7,$ $1/100,$ $1/1000$ for air, water, and liquid He.
The higher viscosity of stellar plasma is more than compensated for by the larger length scales appropriate to stars.
% For most stars $ S_\gamma /{\cal R} \sim T_7^3/\rho $. $v \sim T^{1\over2}$,
%${\cal R}_e \sim $, what is coulomb logarithm?, $Z\sim 1$
In the laboratory, there is a transition from laminar to turbulent flow at about ${\cal R}_e \sim 10^2 $ to $2 \times 10^3$, depending upon the geometry of the boundaries.
We note $\nu$ is essentially the product of a mean-free-path and a microscopic velocity.
Microscopic velocities are of the order of the sound speed, so that macroscopic velocities are usually smaller (but not by much, e.g., Mach numbers $\sim 0.001$ to $0.1$ are typical).
Because the macroscopic length scales characteristic of stars are so much larger than microscopic mean-free-paths ($\sim 10^{18}$, e.g.), ${\cal R}_e \gg 2 \times 10^3$, the flow in stars generally is expected to be highly turbulent (Hansen \& Kawaler 1994, Hansen, Kawaler \& Trimble 2004). In special cases strong rotation and strong magnetic fields may inhibit turbulence.

%	what is a turbulent velocity field? non steady flow, sensitive to initial conditions, separation of trajectories, fluctuations not perturbations, importance of averaging, 3D cascade to Kolmogorov scale, transport dominated by large scale
	
In addition to viscosity (the transport of momentum), thermal diffusivity (heat transport) and molecular diffusivity (composition transport) need to be considered.	
This hierarchy could be summarized by:
\begin{equation}
1 \ll \ell \cdot u /\chi \ll \ell \cdot u/\nu \sim \ell \cdot u/D,
\end{equation}
where $\chi$, $\nu$ and $D$ are the thermal diffusivity, kinematic viscosity and molecular diffusivity, respectively. This order is essentialy determined by the relevant cross-sections for the separate processes;
collisional cross sections for ions determine the kinematic viscosity (ions carry most of the momentum in the plasma) and the molecular diffusivity (ions carry the compositional identity). In contrast, radiative cross-sections for electrons are important for thermal diffusivity. In direct numerical simulations (DNS), the zone size is chosen to be smaller than the smallest relevant length scale (Pope, 2000, Ch.~9).
Given finite computer power, this limits the size of the object that can be simulated. Stars are simply too big to be done with DNS. In this sense stellar simulations are under-resolved.  All simulations of stars are large eddy simulations (LES), with the consequent issue of whether the sub-grid physics is properly represented. For turbulence the sub-grid physics implicit in a set of shock-capturing hydrodynamic algorithms (one of which we use) is in fact appropriate (Boris 2007); such simulations are termed ``implicit LES'' or ILES. This good fortune is due in part to the nature of the dissipation in a turbulent cascade (Kolmogorov 1962), for which 
\begin{equation}
\epsilon = {4 \over 5} v_{rms}^3 / \ell_d \label{eps}
\end{equation}
is the rate that large scale kinetic energy is fed into the cascade. Here $v_{rms}$ is the average turbulent velocity and $\ell_d$ is the characteristic linear dimension of the turbulent region. There is no explicit viscosity; the nonlinear flow adjusts itself to ``hide'' the viscous term in the Navier-Stokes equation, and essentially replaces it by the effective dissipation, $-\epsilon$. The effective viscosity for the turbulent case has little to do with the viscosity for laminar flow.

What about the other processes?
Viallet et al. (2013) show that the transport properties are dominated by large scale coherent motions which are resolved in the body of the convective region. At the boundaries the current resolutions used may be too coarse,  so that the mixing rates seen in the simulations may be affected by the numerical method and the resolution.
There may be other physical processes that could remain completely unresolved in 3D simulations (e.g. rotationally induced shear mixing and MHD), for which discovery of a good algorithm (and correct physics) remains a challenge. 

 Viscosity smooths momentum gradients, so that a finite space-time grid automatically loses detailed information about momentum at the grid scale, and this gives rise to a ``numerical viscosity".
 Computer turbulence simulations have limits on the values of Reynolds numbers that may be simulated; we may define a ``numerical Reynolds number'': $[{\cal R}_e]_{num}= r  u/ \Delta r \Delta u \propto n^{4/3}$, where $n=r/\Delta r$ is the number of zones along a linear dimension $r$ and $\Delta u \sim (\epsilon\ \Delta r)^{1 \over 3}$ is estimated from a Kolmogorov (1962) cascade. For 3D flow, the computational cost is proportional to $n^4$ (spacetime is 4D), so that the difference in cost between $n\sim 100$ and $n\sim1000$ is a factor of 10,000.
For $n\sim 100$  numerical viscosity is significant (think molasses) while for $n \sim 1000$ the flow is clearly turbulent (think water). The difference in computational cost is roughly equivalent to the power of a laptop compared to a supercomputer (2 cores to 20,000 cores).

%	3D sims with enough resolution to be turbulent Meakin-Arnett use sector, Herwig-Woodward larger domains

%	ILES (limitations?)

\subsection{Von Neumann's Dream}

John von Neumann  envisaged using computers to help understand turbulence (von Neumann 1948); this is an example of his idea that numerical experiments, when studied, would inspire mathematical insight, resulting in improved theoretical understanding. Although early 2D simulations (Baz\'an \& Arnett, 1994, 1997a, 1997b, 1998) indicated that 1D stellar evolution models had problems, 2D hydrodynamics can be dangerously unphysical except in special cases in which there is a physical process to enforce the restricted symmetry. The first 3D simulations with the same microphysics as 1D stellar models, and having adequate resolution to exhibit turbulent flow, were those of Meakin \& Arnett (2007), which modeled a sector of an oxygen burning shell during its early stages.  The simulation developed convection quickly, and showed turbulent pulses (about 8 were computed at that time). The convection was not steady-state except in a coarse-grained, average sense.  The average rate of buoyancy work was approximately balanced by the rate of kinetic energy dissipation expected from the turbulent cascade (Kolmogorov 1962), but the instantaneous values of dissipation lagged the driving by a significant fraction of a transit time for the convective zone (Arnett, Meakin \& Young 2009). This dynamic behavior was strikingly similar (see Arnett \& Meakin 2011b) to the famous, and far simpler model of a convective roll due to Lorenz (1963), which is a classic example of deterministic chaos. 
A second striking feature of these simulations was the appearance of entrainment at the boundaries of the convective region, which was quantitatively similar to experimental results (see Meakin \& Arnett 2007) but less like ``overshoot'' prescriptions used in stellar modeling. Vigorous wave generation also occurred at the convective zone boundaries.
% Here is an example of the general form of a figure:
% Fill in the caption in the braces of the \caption{} command. 
% Put the label that you will use with \ref{} command in the braces of the \label{} command.
%
\begin{figure}
\includegraphics[width=4in]{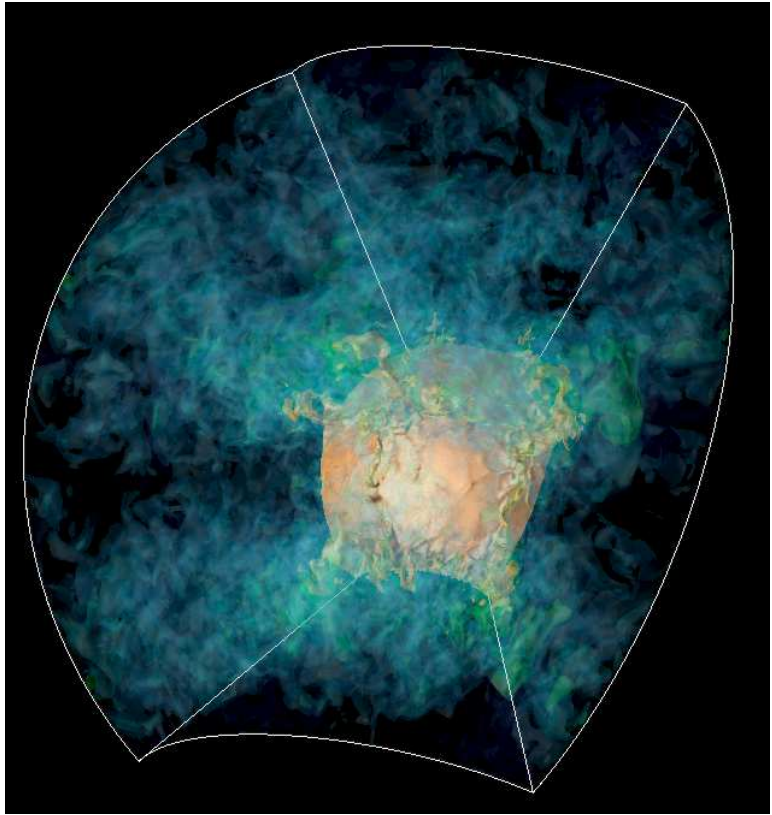}%
\caption{Three-dimensional turbulent mixing in a stratified $\rm ^{16}O$ burning shell which is four pressure scale heights deep. The ashes ($\rm ^{32}S$, yellow) are being dredged up from the underlying core (orange, notice surface waves).  The multi-scale structure of the turbulence is prominent.  Entrained material is not particularly well mixed, but has features which trace the large scale advective flows in the convection zone.  
Also visible are smaller scale features which are generated as the larger features become unstable, breaking apart to become part of the turbulent cascade. The white lines indicate the boundary of the computational domain, which is $120^\circ \times 120^{\circ} $  in angle including the equator, with polar axis aligned from bottom to top. \label{Figure1}}%
\end{figure}

\begin{figure}
\includegraphics[width=2.5in]{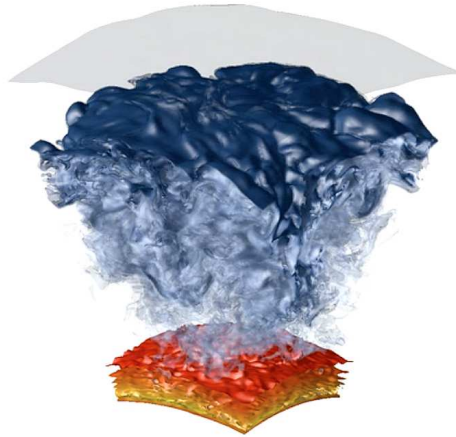}%
\caption{Oxygen burning in a moderately stratified shell (two pressure scale heights deep, enclosed by stably stratified layers. This figure shows isocontours of oxygen mass fraction (blue) as well as nuclear energy generation (red).  The uppermost opaque isocontour (blue) of oxygen mass fraction illustrates the unstable motions in the stably stratified boundary layer as well as showing how this material is dredged into the convection zone in spots. The region of most active burning (yellow) is thin. Burning occurs intermittently, in cells which drive plumes of ascending matter, and is steady only in a crude average sense. \label{Figure2}}%
\end{figure}

Figures~\ref{Figure1} and \ref{Figure2} give a sense of the turbulent behavior. Both simulations were of oxygen burning in a massive star, prior to silicon burning and core collapse, in the last day of the star's life. The energy available in oxygen burning is still sufficient at this stage to blow the star apart, if released explosively. Figure~\ref{Figure1} illustrates turbulent mixing in a stratified burning shell having a depth of four pressure scale heights. The yellow color represents the abundance of $\rm ^{32}S$, which is a product of oxygen burning. It is produced directly by the thermonuclear fusion of $\rm ^{16}O$ in the burning shell, and by erosion of the already burned core (the orange spherical surface) by entrainment. The turbulent region is bounded by stably  stratified fluid and exhibits strongly dynamic wave motions. The $\rm ^{32}S$ abundance traces both large scale structure of advective motion, and smaller scale structure resulting from the instability and break-up of large scale motions into the turbulent cascade. 

Figure~\ref{Figure2} shows an oxygen burning shell with less stratification (two pressure scale heights in depth),  illustrating the burning region in more detail. Again, the convective region is enclosed by stably stratified layers above and below. The blue isocontours show $\rm ^{16}O$ mass fraction, with the color map chosen to indicate the difference in value from top to bottom of the convective zone (clear at the bottom and dark at the top). The large distortions in the uppermost opaque contour of $\rm ^{16}O$ are due to surface waves, and illustrate how the unstable motions affect the stably stratified boundary layer.  Plumes of material are dredged into the convection zone in localized down drafts. The bottom of the convective region is also active. Energy generation is indicated by a red color map. The region of most active burning (yellow) is thin. Below this another stably stratified region which also is buffeted by turbulent flow. Burning occurs intermittently, in cells which drive plumes of ascending matter, and is steady only in an average sense.

The numerical simulations are a powerful tool with which to study stars, but with limitations: it is not feasible at present to calculate a significant fraction of  a stellar lifetime, even for the relatively short-lived massive stars. That takes too many hydrodynamic times, requiring too many time steps. If we can construct a reasonably realistic initial state, we can simulate interesting and rapidly evolving stages of a star's life. We need a better understanding of the convective flow (which we can directly simulate for short times) to develop a theory of stellar convection for long evolutionary times, which goes beyond the currently used MLT with its heuristic diffusion.

\subsection{RANS}
The Reynolds averaged Navier Stokes (RANS) equations provide a different and useful representation of the physics of turbulence.  Turbulence shows vigorous fluctuations, but a robust underlying regularity. The RANS equations spring from the Navier-Stokes equations, which imply conservation of mass, momentum and energy in a fluid (Landau \& Lifshitz 1959). In the RANS approach a variable $v$ is represented in terms of its average value $\langle v \rangle $ and its fluctuating value $v'$, so $v = \langle v \rangle + v'$ and $\langle v' \rangle = 0$.
The separation of fluctuating and average quantities helps explore the underlying regularities. The averaging operation may involve averaging over time and/or space. While the RANS equations represent the full features of turbulence, they are not closed: each set of moment equations (e.g., for a second order moment $\langle v'v'\rangle$) generates a higher order moment (e.g., a third order moment $\langle v'v'v'\rangle$), so that another equation is needed, and so on (Tennekes \& Lumley 1972). The reason behind this lack of closure is the introduction of fluctuations $v'$ in an insufficiently constrained manner (all such fluctuations are not dynamically self-consistent, so some scheme to eliminate the unphysical ones is required). 

An extensive set of mean field equations and data from 3D simulations can be found in Moc{\'a}k et al.(2014).
 
 \subsection{The Turbulent Vortex Approximation}
One theoretical approach to complex problems is to build a simpler, more easily soluble model to capture some of the features of the complete problem. The numerical simulations of turbulence motivate the ``turbulent vortex model'' (Arnett \& Meakin 2011a), both by quantitative analysis and visual inspection.
The basic idea of the vortex model is to separate the  two distinct aspects of turbulence: the large scale flow as exhibited in the Lorenz convective roll, and a dissipation at the much smaller Kolmogorov scale.
The rate at which energy is fed into the turbulent cascade from large to small scales is given by the ``four-fifths law'' of Kolmogorov (Eq.~\ref{eps}), which has been called one of the few exact and fundamental results of turbulence theory (Frisch 1995).
This acts in conjuction with buoyant driving at the large (integral) scale, as described by the Lorenz model of a convective roll. A dynamic system may be constructed by equating the time derivative of the mean turbulent kinetic energy to the difference of this driving and damping. This system describes a convective cell (analogous to a thunderstorm cell), not the whole convection region which is comprised of many such cells. Finding the correct ensemble average of an array of such cells is a challenge at present.
The turbulent vortex model has the advantage that it may be directly compared to some previous suggestions for time dependent convection (e.g., Gough 1976, Arnett 1972, Stellingwerf 1984), and because it defines an acceleration equation, perhaps it may be generalized to include additional physical effects (composition gradients, rotation, and possibly magnetic fields).
 
%	lorenz chaos dynamic systems (Uriel Frisch*)
%	vortex model = lorenz*+kolmogorov* (but cell, building block)
	
\subsection{Closure of the RANS Equations}	
Another, more systematic approach to solving  the turbulence problem is to approximate the complete system by enforcing closure on the Reynolds averaged Navier Stokes (RANS) equations at some level. To be  soluble an approximation must be made to relate the highest order moments with lower order ones, closing the system. While simple in principle, the practice is difficult. The RANS equations are exact, in principle, prior to closure; the approximation used to close them is the issue.
The simulations are well behaved when analysed from the RANS point of view (Viallet, et al., 2013), and promise to further our theoretical understanding, especially if there is an interplay between the more formal and precise RANS approach and the more physically transparent simple modeling.

A similar problem has been an issue in atmospheric science for a long time (e.g., Lumley \& Panofsky 1964), and some success has resulted from closure by modifying the equations of third moments (Lumley, Zeman \& Siess 1978). This is in contrast with MLT used in astrophysics, which only involves modifying the equations of second moments.  Canuto (2011) has decryed this lag by the astrophysical community, and presented a formalism for the stellar case which he terms ``Reynolds stress models"for these long evolutionary times (RSM), which are based on the geophysical studies and have a third order closure. They have not yet been widely applied, possibly because they are more complicated and there is ``no widely accepted stellar evidence that they are superior to MLT for building stellar models''. It now appears, observationally, numerically, and theoretically, that this is {\em not} the case (Meakin \& Arnett 2007b; Smith \& Arnett 2013, and below). It is plausible that asteroseismological studies (e.g., Aerts, et al. 2010, Aerts 2014), which now are showing inadequacies of MLT-built stellar models, may be aided by more sophisticated and independently testable treatments of convection. Although our discussion and that of Canuto might appear at first sight to be different because of their very different origins, they seem to point toward the same sort of improved convection model (compare Canuto 2011 and Viallet, et al., 2013).
However, stars are plasma and geophysical flows are mostly neutral fluids; these differences may become more apparent when magnetic fields are considered.

%The two approaches to approximation are not necessarily independent

%composition and buoyancy,
%	premixed flames (versus p ingestion), composition effects not in MLT

%	tau mechanism (few cells) stochastic wave generation (many cells)
%	Unno*, Cox*, Goldriech*, stochastic generation of waves by convection

%Also "second surprise" entrainment (similar to experiment), less so to astro "overshoot" MA07, MMVA? and extensive wave generation at CZ boundaries.

%SN are major engines for nucleosynthesis

%Some New developments and possible implications: fixing (adding) a missing link and making connections to turbulence and dynamical systems theory

\subsection{\label{sec:level1}Nucleosynthesis}

The nature of this turbulent velocity field $\bf u$ has a direct impact on stellar nucleosynthesis.  
Nuclear reactions conserve baryon number, but not mass. The continuity equation
\begin{equation}
\partial \rho / \partial t + {\bf \nabla \cdot } \rho {\bf u}= 0,
\end{equation}
represents the conservation of a quantity $\rho$ in a relativistically invariant way (a four-vector). We define $\rho = \Sigma_i N_i A_i m_{amu}$ to express baryon number in atomic mass units, where $N_i$ is the volumetric density of species $i$, $A_i$ is the number of baryons in each $i$ nucleus, and $m_{amu} $ is the atomic mass unit. This $\rho$  is numerically close to the ``mass density'', $\rho_m =\Sigma_i N_i m_i$,
where $m_i$ is the mass of each nucleus of the $i$ species, as used in Landau \& Lifshitz (1958) for example. The difference is of order of the nuclear binding energy over $A_i m_{amu}c^2$, summed over all $i$ species; this is small and may be neglected except for extreme conditions (e.g., core collapse). While $\rho$ is exactly conserved and relativistically invariant, $\rho_m$ is not.
Consider the abundance $Y_i = N_i / \rho N_0$ for nucleus $i$, where 
 $N_0= 1/m_{amu}$ is Avogadro's number.  
Using the continuity equation, the co-moving time derivative in the rest frame may be written as
\begin{equation}
dY_i /dt = {\rm nuclear\  reactions} + {\rm microscopic\ diffusion}. \label{sources}
\end{equation}
The ``diffusion'' term is small for massive stars and may be neglected  with regard to its action on the mean fields (however, see Michaud, et al., 2004; Thoul, et al., 1994), but is essential for mixing material to the molecular scale in the turbulent convection zones. The ``reaction'' term is actually comprised of [the sum of all reactions which create nucleus $i$] minus [the sum of all reactions which destroy nucleus $i$]; this is a reaction {\em network} (Arnett, 1996, Ch.~4).
In a given zone in a stellar model, the time derivative of the abundance is
\begin{equation}
d Y_i / dt = \partial Y_i /\partial t + {\bf u \cdot \nabla} Y_i. \label{DYi}
\end{equation}
If the rate of mixing is rapid relative to the rate of nuclear reactions, the gradient of $Y_i$ is zero and the second term may be neglected. 
Using the continuity equation, Eq.~\ref{DYi}  and  \ref{sources} may be written as a conservation equation for $Y_i$, with source and sink terms for nuclear reactions creating and destroying species $i$ on the right hand side, and the divergence of a composition flux replacing the spatial derivative term, where the composition flux is
\begin{equation}
 {\bf F_i} = \rho Y_i {\bf u}.
\end{equation}
Notice that this definition of composition flux does not involve a spatial derivative.
%Historical approach (nuclear, agwc, b2fh, cfz, ann rev, wda96* and processes) used a co-moving derivative and assumed a simple flow (ignored the ${\bf u \cdot \nabla} Y_i $ term).
%\footnote{While most nuclear reaction rates are now reasonably well known,c12(a,g) thorn left.} 
An expression for the mean rate of change of composition at a given stellar radius (as opposed to the instantaneous and local expressions of eq. 4 and 5) can be developed after splitting fields into means and fluctuations, and we find
\begin{align}
\erho\favgDt \favg{Y_i} = -\nabla\cdot \favg{\mathbf F}_{i} + \erho~\favg{\dot{Y}}_{i,{\rm nuc}}
\end{align}
\noindent where the mean Lagrangian derivative is
\begin{align}
\favgDt{Y_i} = \partial_t \favg{Y}_i + \favg{{\mathbf u}}\cdot\nabla\favg{Y}_i \label{DYibar} \end{align} 
\noindent  the turbulent flux of composition of isotope $i$ is
\begin{align}
\favg{\mathbf F}_i = \erho \favg{ \ff{Y}_i \ff{\mathbf u}} \end{align}
\noindent and the tilde and double primes indicate the Favrian mean and fluctuation, respectively (see Viallet, \etal 2013). 

An important issue now arises: while we may equate Eq.~\ref{DYi} to local reaction source terms, 
Eq.~\ref{DYibar} requires reaction source terms which account for both mean and fluctuating components of $Y_i$. Stellar codes almost always use the mean value $\favg{Y}_i$ and ignore contributions due to fluctuations. This is valid when the effects of the fluctuations average to zero, or when the fluctuations are small, as might be expected for fast mixing (turnover time much smaller than reaction time).
The late stages of stellar evolution are replete with fast reactions which come into balance with other reactions tending to cancel their effect. Mixing can affect this balance. While convective turnover may be ``fast'', there will be some reactions which are faster as well as some which are slower. The fast reactions are local while the slow reactions are a global feature of the convection zone. The net effect of this needs to be explored more deeply, perhaps on a case by case basis. Proton ingestion into He burning regions 
(see~\ref{conv_bnd}) and the Urca process (e.g., Arnett 1996) are interesting examples in which this might be significant.

At present it is more common in stellar evolution codes  to replace the turbulent flux term $\favg{\mathbf F}_i$ 
by a heuristic operator for the divergence of ``convective diffusion'' (e.g., Eggleton 1972, Weaver, Zimmerman, \& Woosley 1978, Langer, El Eid, Fricke 1985, Langer 2012), 
\begin{equation}
 - {\partial \over \partial m }\Big [ (4 \pi r^2 \rho)^2D {\partial Y_i \over \partial m }\Big], 
\end{equation}
where $ D = { 1 \over 3} \ell\, u_{rms} $ with $\ell$ being the mixing length and $u_{rms}$ being the rms value of the turbulent velocity as estimated from MLT; this
smooths composition gradients by mixing. 
This is equivalent to replacing the radial component of the composition flux by 
\begin{equation}
F_i = - \rho D {\partial Y_i \over \partial r}, 
\end{equation}
and setting the other components to zero.
This procedure is dangerous from a mathematical standpoint because it increases the order of the spatial derivative by one, and thus introduces spurious additional solutions. It is also dangerous from a physical standpoint because it makes the composition currents large in composition gradients (formally infinite at a contact discontinuity in composition; this must be avoided by setting $u_{rms}$ to zero appropriately).
In contrast, the numerically simulated composition currents are large in the interior of the convection zone, and smaller at the boundaries, a qualitatively different behavior.
Because this approach to mixing is heuristic, it gives no help with the issue of the effect of fluctuations in composition in a turbulent burning region, and in addition, could be misleading with regard to estimated rates of mixing and burning. This is especially worrisome for the late stages, e.g., prior to core collapse in massive stars, which are more poorly constrained by observational data.

\section{Implications for Astrophysics}

Why replace MLT with a more complex algorithm? Because there are phenomena which MLT does not describe correctly, and that have observational consequences which may already have been detected. Also, a classical problem in stellar pulsation theory is the role of convection in driving pulsations, which is ignored, as lamented by Unno et al. (1989), but seems to be related to chaotic convection (Arnett \& Meakin 2011b).

\subsection{Convective Boundaries\label{conv_bnd}}
B\"ohm-Vitense (1958) considered a well-mixed plasma, with uniform composition, so that MLT has no inherent ability to deal with nonuniform composition. In practice stellar evolution codes use the Schwarzschild criterion (independent of composition) or the Ledoux criterion (composition dependent), and sometimes  a combination of both; the simulations do not seem to validate this picture. The problem is simply illustrated. Convective flow at low mach number may be regarded as buoyancy driven, where the buoyant acceleration is $g  \rho' / \rho$. For an ideal gas at constant pressure (i.e., low Mach number flow), $\rho'/\rho +T'/T + Y'/Y=0$, so that fluctuations in composition (which are ignored) come in at the same level as temperature fluctuations (which are considered). Further, in a turbulent medium, composition fluctuations are homogenized, so the convective boundary is a special place, having composition gradients on one side and homogeneity on the other. The simulations show that such regions have complicated dynamic behavior which may not yet be resolved (see however, Woodward et al., 2014), but give interesting suggestions to compare with asteroseismological data (e.g.,  see the discussion of the ``overshoot'' parameter in Aerts 2014). 
This physics is also relevant to proton ingestion and nucleosynthesis in the AGB shell flash (Herwig 2000, Stancliffe et al. 2011,Woodward et al., 2014), and in the He core flash (Mo\'cak, et al., 2011).

\subsection{Presupernova Eruptions}
There is a considerable body of observational evidence that many massive stars have eruptive episodes prior to core collapse (Smith \& Arnett 2013). Most obvious are the supernovae of types In and IIn, which require mass ejection just prior to core collapse in order to provide the mass of material which is seen to interact with the supernova ejecta. 
It has been shown that MLT is a steady state approximation, and that if this restriction is relaxed the convection in late evolution of massive stars is strongly dynamic and chaotic (Smith \& Arnett 2013, Arnett \& Meakin 2011b, Meakin \& Arnett 2007b). This may lead to mass loss, either steady but vigorous (Quataert \& Shiode 2012, Shiode \& Quataert  2013) or eruptive, as observations indicate.

\subsection{Realistic Progenitors for Gravitational Collapse}
Gravitational collapse of stellar cores to form neutron stars or black holes is an initial value problem, so that it is important to have realistic initial values, i.e., realistic progenitor star models.   The first multidimensional simulation of a massive star for a significant length of time was done by Meakin (2006); earlier work by Baz\`an \& Arnett (1994, 1997, 1997b, 1998) was limited to shorter intervals of time by lack of computer power.
The results were so startling that extensive checking was required prior to publication (Arnett \& Meakin 2011a). In these 2D simulations, which included active shells of C, Ne, O and Si burning, the static initial models began to convect, the flow became steadily more vigorous, building radial pulsations, and finally pushing so much matter off the grid that the simulations were terminated.
Murphy et al. (2004) had correctly shown that the nuclear burning was relatively stable to the $\epsilon$-instability. Instead,
this instability was driven by the time-dependent turbulent convection ($\tau$-instability, Arnett \& Meakin 2011b).
All this action occurred near, but prior to, core collapse, and modified the progenitor significantly from the original 1D stellar model. Because 2D simulations may be misleading for 3D problems, these simulations need to be done in 3D, with a larger domain and a full $4\pi$ steradian grid. Until such simulations have been done and carefully analyzed, the detailed predictions of 1D stellar evolution codes for core collapse progenitors should be viewed with serious caution.

Couch \& Ott (2013) have shown that qualitative changes may result in simulations of neutrino driven explosions; they carried out a controlled set of simulations of the core collapse of a 15 \sol\ dud of Woosley \& Heger (2007). Simply mapping a velocity structure similar to the convective rolls of Meakin \& Arnett (2007b) and Arnett \& Meakin (2011a), and slightly increasing the neutrino heating (5 percent) changed a dud into a successful explosion. Without the changed velocity structure but still with the slightly enhanced neutrino heating, there was no explosion.  Couch \& Ott (2013) conclude that ``initial conditions matter'' for the core collapse problem.
It is interesting that the same turbulence physics which is important for progenitors is also useful in understanding the behavior of 3D collapse simulations themselves (Murphy  \& Meakin 2011).
We may ask to what extent are the theoretical difficulties with core collapse supernova models due to incorrect progenitor (``initial'') models?

\subsection{Other Implications and Issues}
{\bf Meteoritic Anomalies.} The meteoritic abundances, especially isotopic ratios, of carbonaceous chondritic meteorites have been a key empirical test for theories of nucleosynthesis. For example,
Amari, \etal (2012) use quantitative details of 1D stellar models for estimates of nucleosynthesis yields.  Simulations show that rather than one abundance pattern there are several clustered around a mean in a turbulent convection zone (Meakin \& Arnett 2007b). Beyond this there will be systematic shifts due to the difference between 3D models and 1D models. There will be ``extra mixing'' because of inadequate treatment of boundaries in 1D. Advection can give large scale transport without microscopic mixing, especially during violently dynamic evolution, so ``layering'' can be violated. Some of the existing puzzles in interpretation  of meteoritic anomalies in abundance may be due to over-simplified and unphysical mixing in 1D models.
%	Yields, extra mixing, Òjumping layersÓ in meteorites

{\bf Novae.} Kelly \etal (2013) have suggested that classical novae may be used as indicators of mixing, and identify key abundance ratios which may be most useful. This analysis is entirely based on a 1D framework of initial and hydrodynamic models. This appears to be a promising approach, but needs to be expanded to include effects discussed above (3D is not 1D); 3D effects in (1) the initial hydrostatic model, (2) the build up to instability and (3) the hydrodynamic response may affect the predictions of nucleosynthesis.

{\bf Rotation.} A fundamental problem in stellar evolution is the effect of rotation (Maeder \& Meynet 2000); stars are observed to have rotation rates ranging from slow to near breakup. Stars are plasma so that rotation should not be discussed without dealing with magnetic fields (the dynamo problem), as the solar surface and pulsars gently remind us. It is thought that gamma-ray bursts (GRB's) are the consequence of rotating core collapse. The best attempts to model the progenitors of such events (e.g., Groh, \etal 2013) use a diffusion model of convection, which  seems unlikely to be correct. There is a dimensionless ratio, the Rossby number (ratio of convective velocity to rotational velocity), which separates qualitatively different types of flow. This has not yet been explored in the stellar case for high numerical Reynolds numbers and realistic boundary conditions; see Miesch (2005) for a review of simulations at moderate Reynolds numbers, and Woodward, Herwig \& Lin (2014) for the current state of the art.  

%	rotation and GRBÕs still to do, acceleration equation or RANS, Maeder \& Meynet

{\bf The Sun.} The Sun may provide clues. Goldreich \etal (1994) showed the importance of solar convection in driving ``solar-like'' oscillations, which are key to helioseismology, and asteroseismology of stars similar to the sun.
Helioseismology provides precise measures of the convection zone and the tachocline. Recent work by Balbus (2009) seems to suggest that a simple extension of the turbulent vortex model may be possible, using the ``wind equation'', which would describe the differential rotation within the Sun.
This seems to be a prudent first step to complete as we progress toward understanding GRB's.

%	SSM get CZ and tachocline, differential rotation and Balbus, Òwind equationÓ, 		
%	Goldreich-Kumar-Murray? driving Òsolar like oscillationsÓ

 \section{Conclusions}
 Moore's law (see e.g., Gottbrath et al. 1999) has brought us to the point where present day computer power allows ILES simulations with sufficient resolution to give turbulent flow in 3D. We find new phenomena not predicted by standard stellar evolution codes, which use MLT. Theoretical analysis of the simulations promises a deeper understanding of nucleosynthesis, and of the dynamics of the late stages of evolution of massive stars (prior to core collapse and possible supernova explosion), as well as of mixing and burning in stars which are asteroseismological targets.

\acknowledgements{This work was supported in part by NSF Award 1107445  at the University of Arizona. One of us (DA) wishes to thank  the Aspen Center for Physics (ACP) and the Kavli Institute for Theoretical Physics (KITP) for their hospitality and support.}

%\appendix

%\section{Appendixes}

\section{References}

%\begin{thebibliography}
\begin{itemize}
%
%\item []{}Abramowitz, M., \& Stegun, I. A., 1964,
%{\it Handbook of Mathematical Functions}, National Bureau of Standards, Applied
%Mathematics Series 55

%%\bibitem[Abramowitz \& Stegun(1964)]{abram} Abramowitz, M., \& Stegun, I. A., 1964,
%%{\it Handbook of Mathematical Functions}, National Bureau of Standards, Applied
%%Mathematics Series 55
%
\item[]{} Aerts, D., 2014, in {\it Setting a New Standard in the Analysis of Binary Stars}, ed. K Pavlovski, A. Tkachenko and G. Torres, EAS Publications Series, and arXiv:1311.6242v1

\item[]{} Aerts, C., Christensen-Dalsgaard, J., \& Kurtz, D. W., 2010, {\it Asteroseismology}, Springer, Dordrecht

%%\bibitem[Alligood, Sauer \& Yorke(1996)]{asy96} Alligood, K. T., Sauer, T. D.,
%%\& Yorke, J. A., 1996, {\it Chaos: An Introduction to Dynamical Systems},
%%Springer, Berlin
%
\item[]{} Amari, S., Zinner, E., \& Gallino, R., 2012, {\it 43rd Lunar and Planetary Science Conf.}, 43, 1031

%%\bibitem[Andersen, Clausen, Magain(1989)]{and89} Andersen, J.,
%%Clausen, J. V., \& Magain, P. 1989, \aap, 211, 346
%%\bibitem[Arnett(1994)]{da94} Arnett, D., 1994, \apj, 427, 932
%
%%\bibitem[Arnett(1968)]{wda68} Arnett, W. D., 1968, GISS conference on Supernovae, ed. P. Brancazio and A. G. W. Cameron, Gordon and Breach,
%%New York; also Orange-Aid preprint OAP-132
%
%\item []{} Arnett, W. D., 1969, \apss, 5, 180

\item[]{} Arnett, D., 1972, \apj, 176, 681 %(helium cores)
%

%\item[]{} Arnett, W. D., 1972, \apj, 173, 393
%
%%\bibitem[Arnett(1972c)]{wda72c} Arnett, W. D., 1972, \apj, 176, 699
%
%%\bibitem[Arnett(1972d)]{wda72d} Arnett, W. D., 1972, \apj, 178, 771
%
%%\bibitem[Arnett(1973)]{wda73} Arnett, W. D., 1973, \apj, 179, 249
%
%%\bibitem[Arnett(1974a)]{wda74a} Arnett, W. D., 1974, \apj, 193, 169
%
%%\bibitem[Arnett(1974b)]{wda74b} Arnett, W. D., 1974, \apj, 194, 373
%
%%\bibitem[Arnett(1977)]{wda77} Arnett, W. D., 1977, \apjs, 35, 145
%
\item[]{} Arnett, W. D., 1977, \apj, 218, 815 %(neutrino trapping)
%
%%\bibitem[Arnett(1994)]{wda94} Arnett, D., 1994, \apj, 427, 932
%
\item []{}Arnett, D., 1996, {\it Supernovae and
Nucleosynthesis}, Princeton University Press, Princeton NJ
%%
%% \bibitem[Arnett \& Meakin(2009)]{am09iau} Arnett, W. D., \& Meakin, C.,
%% 2009, in {\it Chemical Abundances in the Universe: Connecting First Stars to
%% Planets}, IAU Symposium 265, ed. K. Cunha, M. Spite, \& B. Barbuy,  
%% 
\item[]{} Arnett, W. D., Meakin, C.,
 \& Young, P. A., 2009, \apj, 690, 1715 %(velocity field)
%% 
%\item[]{} Arnett, W. D., Meakin, C.,
 %\& Young, P. A., 2010, \apj, 710, 1619 %(subphotospheric)
%%
%%\bibitem[Arnett \& Meakin(2011a)]{am11a} Arnett, D., \& Meakin, C., 2011a,
%%{\it Astrophysical Dynamics: From Galaxies to Stars}, IAUS 271,
%%ed. N. Brummell \& S. Brun.
%%
\item[]{} Arnett, D., \& Meakin, C., 2011a,
\apj, 733, 78 %(realistic progenitors)
\item[]{} Arnett, D., \& Meakin, C., 2011b,
\apj, 741, 33 %(turbulent cells)
%
%%%\bibitem[Arnett, Bahcall, Kirshner \& Woosley(1989)]{abkw89} Arnett, W. D.,
%%%Bahcall, J. N., Kirshner, R. P., \& Woosley, S. E., 1989, \araa, 27, 629
%%
%%\bibitem[Arnett \& Ott(2012)]{ao12} Arnett, W., D., \& Ott, C., 2012, \apj,
%%submitted
%%
%%%\bibitem[Asida(2000)]{asi00} Asida, S. M. 2000, \apj, 528, 896
%%
%\bibitem[Asida \& Arnett(2000)]{aa00} Asida, S. M., \& Arnett, D., 2000,
%\apj, 545, 435
%%
%%%\bibitem[Asplund(2005)]{asp05} Asplund, M., 2005, \araa, 43, 481
%%
\item[]{} Balbus, S. A., 2009, \mnras, 395, 2056
%%
%%\bibitem[Balbus \& Weiss(2010)]{bw10} Balbus, S. A. \& Weiss, N. O., 2010,
%%\mnras, 404, 1263
%%
%%%\bibitem[Balbus \& Hawley(1994)]{bh94} Balbus, S. A., \& Hawley, J. 
%%%F., 1994, \mnras, 266, 769
%%
%%%\bibitem[Balbus \& Hawley(1998)]{balbus} Balbus, S. A., \& Hawley, J. 
%%%F., 1998, Rev. Mod. Phys., 70, 1
%%
%%%\bibitem[Barrado y Navascu\'{e}s, Stauffer, \& Patten(1999)]{nsp99}
%%%Barrado y Navascu\'{e}s, D., Stauffer, J. R., \& Patten, B. M. 1999,
%%%\apj, 522, L53
%%
%%%\bibitem[Basu \& Antia(1997)]{ba97} Basu, S., \& Antia, H. M., \mnras, 287, 189
%%
%%%\bibitem[Basu \& Antia(2004)]{ba04} Basu, S., \& Antia, H. M., \apj, 606, L85
%%
%%%\bibitem[Basu, Pinsonneault, \& Bahcall(2000)]{bpb00} Basu, S.,
%%%Pinsonneault, M. H., \& Bahcall, J. N., 2000, \apj, 529, 1084
%%
\item[]{} Baz\`{a}n, G., \& Arnett, D., 1994, \apj, 433, L41
\item[]{} Baz\`{a}n, G., \& Arnett, D., 1997, Science, 276, 1359
\item[]{} Baz\`{a}n, G., \& Arnett, D., 1997b, Nucl. Phys. A, 621, 607
\item[]{} Baz\`{a}n, G., \& Arnett, D. 1998, \apj, 494, 316
%%
%%%\bibitem[Berger, \etal(2010)]{berger10}Berger, T. E., Slater, G., Hurlburt, N., Shine, R.,
%%%Tarbel, T., Title, A., Lites, B. W., Okamoto, R. J., Ichimoto, K., Katsukawa, Y., Magara, T.,
%%%Suematsu, Y., Shimizu, T.,  2010, \apj, 716, 1288
%%
%%%\bibitem[Bethe(1990)] {bethe90} Bethe, H. A., 1990, Rev. Mod. Phys., 62, 801
%%
%%%\bibitem[Biermann(1951)]{bier51} Biermann, L., 1951, \zap, 28, 305
%%
%%%\bibitem[Blanco et al.(1980)]{bl80} Blanco, V. M., Blanco, B. M., \&
%%%McCarthy, M. F. 1980, \apj, 242, 938
%%
%%%\bibitem[Bl\"{o}cker(1995)]{blo95} Bl\"{o}cker, T. 1995, \aap, 297, 727
%%
%%%\bibitem[Bodansky, Clayton \& Fowler(1968)]{bcf68} Bodansky, D., Clayton,
%%%D. D., \& Fowler, W. A., 1968, \apjs, 16, 299
%%
\item[]{} Boris, J., 2007,
in {\it Implicit Large Eddy Simulations}, ed. F. F. Grinstein, L. G. Margolin,
\& W. J. Rider, Cambridge University Press, p. 9
%%
%%%\bibitem[Boesgaard \& Friel(1990)]{bf90} Boesgaard, A. M. \& Friel, E. D.
%%%1990, \apj, 351, 467
%%
%%%\bibitem[Boesgaard \& King(2002)]{bo02} Boesgaard, A. M. \& King, J. R.
%%%2002, \apj, 565, 587
%%
\item[]{} B\"ohm-Vitense, E., 1958, \zap, 46, 108
%%
%%%\bibitem[Boothroyd \& Sackmann(2003)]{bs03} Boothroyd, A. I. \&
%%%Sackmann, I.-J. 2003, \apj, 583, 1004
%%
\item[]{} Braginskii, S. I., 1958, {\it Soviet Physics JETP}, 6, 358 

\item[]{} Canuto, V. M., 2011, \aanda,  528, A76
\item[]{} Couch, S. \& Ott, C., 2013, \apjl

%%%\bibitem[Cowling(1941)]{cowl41} Cowling, T. G., 1941, \mnras, 101, 367
%%
%\item[]{} Cox, J. P., 1980, {\it Theory of Stellar
%Pulsations}, Princeton University Press, Princeton NJ
%%
%%%\bibitem[Cox(1968)]{cox68} Cox, J. P., 1968, {\it Principles of Stellar
%%%Structure}, in two volumes, Gordon \& Breach, New York
%%
%%%\bibitem[Cvitanovi\'c(1989)]{cvit} Cvitanovi\'c, P., {\it Universality
%%%in Chaos}, Adam Hilger, Bristol and New York
%%
\item[]{} Daly, B. J. \&  Harlow, F. H., 1970, \physflu, 13, 2634

\item[]{} Eggleton, P. P., 1972, \mnras, 156, 361
%%
%%\bibitem[Fernando(1991)]{fernando} Fernando, H. J. S., 1991, Ann. Rev. Fluid Mech., 23, 445
%%
%%%\bibitem[Fesen, Becker \& Goodrich(1988)] {fesen88} Fesen, R. A., Becker, R. H.,
%%%Goodrich, R. W., 1988, \apj, 329, L89
%%
%%%\bibitem[Feuchtinger, Buchler, \& Koll\'{a}th(2000)]{fbk00}
%%%Feuchtinger, M., Buchler, J. R., \& Koll\'{a}th, Z. 200, \apj, 544,
%%%1056
%%
%%%\bibitem[Fontaine \etal(1981)]{fvw81} Fontaine, G., Villeneuve, F.,
%%%\& Wilson, J., 1981, \apj, 243, 550
%%
%%%\bibitem[Franceschini \& Tebaldi(1979)]{ft79} Franceschini, V. \&
%%%Tebaldi, C., 1979, Journal of Statistical Physics, 21, 707
%%
%\bibitem[Freytag, Ludwig, \& Steffan(1996)]{fls96} Freytag, B.,
%Ludwig, H.-G., \& Steffan, M., 1996, \aap, 313, 497
%%
%%\bibitem[Freytag, et al.(2012)]{co5bold} Freytag, B., Steffan, M., Ludwing, H.-G., Wedemeyer-B\"ohm, W., Schaffenberger, Steiner, O., 2012, J. Comp. Phys., 231, 919, 
%%arXiv:1110.6844v1
%%
\item[]{} Frisch, U., 1995, {\it Turbulence}, Cambridge University Press, Cambridge
\item[]{} Goldreich, P., Murray, N., \& Kumar, P., 1994, \apj, 424, 466
\item[]{} Gottbrath, C., Bailin, 
J., Meakin, C., Thompson, T., \& Charfman, J.~J.\ 1999, arXiv:astro-ph/9912202 
\item[]{} Gough, D., 1976, \apj, 214, 196
%%%\bibitem[Gough(1991)]{gough91} Gough, D. O., 1991, Ann. N. Y. Acad. Sci.,
%%%647, 199
%%
%%%\bibitem[Grevesse \& Sauval(1998)]{gn93} Grevesse, N. \& Sauval,
%%%A. J., 1998, Space Science Reviews, 85, 161
%%
\item[]{} Groh, J. H., Meynet, G., Georgy, C. \& Ekstr\"om, 2013, \aanda, 558, 131
%%%\bibitem[Guzik, etal(2005)]{gwc05} Guzik, J. A., Watson, L. S., 
%%%\& Cox, A. N., 2005, \apj, 627, 1049
%%
\item[]{} Hansen, C. J., \& Kawaler, S. D., 1994, {\it Stellar Interiors}, Springer-Verlag
\item[]{} Hansen, C. J.,  Kawaler, S. D. \& Trimble, V., 2004, {\it Stellar Interiors}, Second Edition, Springer-Verlag
%%
%%%\bibitem[Heger, Langer, \& Woosley(2000)]{hlw00} Heger, A., Langer, 
%%%N., \& Woosley, S. E., 2000, \apj, 528, 368
\item[]{} Henyey, L. G., Wilets, L., B\"ohm, K. H., Lelevier, R., Levee, R. D., 1959, \apj, 129, 628

\item[]{} Herwig, F., 2000, \aanda, 360, 952
%%\bibitem[Herwig, et al.(2011)]{herwig11} Herwig, F., Pignatari, M., Woodward, P. R.,
%%Porter, D. H., Rokefeller, G., Fryer, C. L., Bennett, M., \& Hirschi, R., 2011,
%%\apj, 727, 89
%%
%%%\bibitem[Hillenbrand \& White(2004)]{hw04} Hillenbrand, L. A. \&
%%%White, R. J. 2004, \apj, 604, 741
%%
%%\bibitem[Holmes, Lumley, \& Berkooz(1996)]{hlb96} Holmes, P., Lumley, J. L.,
%%\& Berkooz, G., 1996, {\it Turbulence, Coherent Structures, Dynamical Systems,
%%and Symmetry}, Cambridge University Press
%%
%%%\bibitem[Hoyle(1946)]{fh46} Hoyle, F., 1946, \mnras, 106, 343
%%
%%%\bibitem[Hoyle(1954)]{fh54} Hoyle, F., 1954, \apjs, 1, 121
%%
\item[]{} Hoyle, F., 1955, {\it Frontiers of Astronomy}, 
Harper \& Brothers, New York
%%
%%%\bibitem[Hughes, \etal(2000)]{hughes00} Hughes, J. P., Rakowski, C. E.,
%%%Burrows, D. N., \& Slane, P. O., \apj, 528, 109
%%
%%\bibitem[Hurlburt, Toomre, \& Massaguer(1996)]{hurl96} Hurlburt, N. E.,
%%Toomre, J., \& Massaguer, J. M., \apj, 311, 563
%%
\item[]{} Iben, Icko, Jr., 1965, \apj, 141, 993

\item[]{} Iben, Icko, Jr., 2013, {\it Stellar Evolution Physics: Vol. 1: Physical Proceesses in Stellar Interiors},
Cambridge University Press

%%%\bibitem[Iglesias \& Rogers(1996)]{ir96} Iglesias, C. \& Rogers,
%%%F. J. 1996, \apj, 464, 943
%%
%%%\bibitem[Josselin \& Plez(2007)]{jp07} Josselin, E., \& Plez, B.,
%%%2007, \aap, in press
%%
%%%\bibitem[Kadanoff(1966)]{leo} Kadanoff, L. G., 1966, Physics, 2, 263
%%
%%
%%%\bibitem[Kalogera, Valsecchi \& Willems(2008)]{kvw08} Kalogera, V.,
%%%Valsecchi, F., \& Willems, B., 2008, in {\it 40 Years of Pulsars, Magnetars,
%%%and More}, AIP Conference Proceedings, 983, 433.
%%%
\item []{}Kelly, K. J., Iliadis, C., Downen, L, Jos\`e, J., Champagne, A., 2013, arXiv:1311.2813v1

%%%\bibitem[Kifonidis, \etal(2003)]{kif03} Kifonidis, K., Plewa, T., Janka, H.-Th.,
%%%M\"uller, E., 2003, \aap, 408, 621
%%%
%%%\bibitem[Kifonidis, \etal(2006)]{kif06} Kifonidis, K., Plewa, T., Scheck, L., Janka, H.-Th., M\"uller, E., 2006, \aap, 453, 661
%%
%%\bibitem[Kim et al.(1996)]{kim96} Kim, Y.-C., Fox, P. A., Demarque, P., 
%%\& Sofia, S., 1996, \apj, 461, 499
%%
%%\bibitem[Kim et al.(1995)]{kim95} Kim, Y.-C., Fox, P. A., Sofia, S.,
%%\& Demarque, P.,  1995, \apj, 442, 422
%%
\item[]{} Kippenhahn, R. \& Weigert, A. 1990, {\it Stellar Structure and Evolution}, Springer-Verlag
%%
%%%\bibitem[Kiss, Szabo, \& Bedding(2006)]{kiss06} Kiss, L. L., Szabo, Gy. M., \&
%%%Bedding, T. R., 2006, \mnras, 372, 1721
%%
%\item[]{} Kolmogorov, A. N., 1941, Dokl.  Akad. Nauk SSSR, 30, 299
%%
\item[]{} Kolmogorov, A. N.,1962, J. Fluid Mech., 13, 82
%%
%%%\bibitem[Kraichnan(1987)]{kraichnan} Kraichnan, R. H., 1987, Phys. Fluids,
%%%30, 2400
%%
%%%\bibitem[Kritsuk, \etal(2007)]{knpw07} Kritsuk, A. G., Norman, M. L.,
%%%Padoan, P., \& Wagner, R., 2007, \apj, in press
%%
%%%\bibitem[Kudritzki et al.(1989)]{kud89}Kudritzki, R. P., Pauldrach, A.,
%%%  Puls, J., Abbott, \& D. C.  1989 \aap, 219, 205
%%
%%%\bibitem[Kumar \& Quataert(1997)]{kq97} Kumar, P., \& Quataert, E. J., 1997,
%%%\apj, 475, L143
%%
%%%\bibitem[Kuhlen, Woosley, \& Glatzmaier(2003)]{kwg03} Kuhlen, M.,
%%%Woosley, S. E., \& Glatzmaier, G., {\it 3D Stellar Evolution}, ed.,
%%%Turcotte, S., Keller, S. C., \& Cavallo, R. M., A.S.P. Conf. Series 293
%%%
%%%\bibitem[Lamers \& Nugis(2003)]{ln03} Lamers, H. J. G. L. M. \& Nugis,
%%%T. 2003 \aap, 395L, 1
%%
%%%\bibitem[Landau(1944)]{landau44} Landau, L. D., 
%%%C. R. (Dokl.) Acad. Sci. URSS 44, 311
%%
\item[]{} Landau, L. D. \& Lifshitz, E. M.
1959, Fluid Mechanics, Pergamon Press, London

\item[]{} Langer, N., El Eid, M., \& Fricke, K., 1985, \aanda, 145, 179
\item[]{} Langer, N., 2012, \araa, 50, 107
%%%\bibitem[Lastennet \& Valls-Gabaud(2002)]{las02} Latennet, E. \&
%%%Valls-Gabaud, D. 2002, \aap, 396, 551
%%
%%%\bibitem[Lastennet et al.(1999)]{las99} Lastennet, E., Lejeune, T.,
%%%Westera, P., \& Buser, R. 1999, \aap, 341, 857
%%
%%%\bibitem[Latham et al.(1996)]{lat96} Latham, D. W., Nordstr\"{o}m, B.,
%%%Andersen, J., Torres, G., Stefanik, R. P., Thaller, M., \& Bester,
%%%M. J. 1996, \aap, 314, 864
%%
%%%\bibitem[Ledoux(1941)]{ledoux41} Ledoux, P., 1941, \apj, 94, 537
%%%
%%%\bibitem[Ledoux(1958)]{ledoux58} Ledoux, P., \& Walraven, Th., 1958,
%%%in {\it Handbuch der Physik}, 51, ed. S. Flugge, (Springer-Verlag, Berlin), p. 353
%%
%%%\bibitem[Ledoux(1974)]{doux74} Ledoux, P. 1974, in {\it IAU Symposium
%%%59, Stellar Instability and Evolution}, ed. P. Ledoux et
%%%al. (Dordrecht: Reidel)
%%
%%%
%%%\bibitem[Libchaber \& Mauer(1982)]{libchaber} Libchaber, A., \& 
%%%Mauer, J., 1982, in {\it Nonlinear Phenomena at Phase Transitions
%%%and Instabilities}, ed. T. Riste, 259-86, Plenum Publ. Corp.
%%
%%%\bibitem[Lighthill(1978)]{lh78} Lighthill, J. 1978, {\it Waves in
%%%Fluids} (Cambridge: Cambridge University Press)
%%
%%%\bibitem[Lindl(1998)]{lindl} Lindl, J. D., 1998, {\it Inertial Confinement Fusion},
%%%Springer-Verlag, New York
%%
%%%\bibitem[Livne(1993)]{eli93} Livne, E., 1993, \apj, 406, 17
%%
%%%\bibitem[Lodders(2003)]{lo03} Lodders, K., 2003, \apj, 591, 1220
%%
\item[]{} Lorenz, E. N., 1963, Journal of Atmospheric Sciences, 20, 130
%%
%\item[]{} Lorenz, E. N., 1995, {\it The Essence of Chaos}, University of Washington Press, Seattle
%%
%%\bibitem[Ludwig \etal(1999)]{lfs99} Ludwig, H.-G., Freytag, B., \&
%%Steffan, M., 1999, \aap, 346, 111
%%
\item[]{} Lumley, J. L., \& Panofsky, H. A., 1964, {\it The Structure of Atmospheric Turbulence}, Interscience Publishers, New York
\item[]{} Lumley, J. L., Zeman, O., \& Siess, J., 1978, \jfm, 84, 581
%%%\bibitem[Maeder \& Meynet(1989)]{mm89} Maeder, A. \& Meynet, G. 1989,
%%%\aap, 84, L89
%%
\item[]{} Maeder, A. \& Meynet, G., 2000, \araa, 38, 143
%%
%%
%\item []{} Markova, N., Puls, J., Sim\'on-D\'iaz, S., Herrero, A., Markov, H., Langer, N., 2013, arXiv:1310.8546v1, \aanda, in press

%%
%%%\bibitem[Mamajek(2002)]{eem} Mamajek, E. E. 2002, private communication
%%
%%%\bibitem[Mamajek, Lawson, \& Feigelson(2000)]{mlf00} Mamajek, E. E.,
%%%Lawson, W. A., \& Feigelson, E. D. 2000, \aj, 112, 276
%%
%%%\bibitem[May(1976)]{may} May, Robert M., 1976, Nature, 261, 459
%%

%%%\bibitem[McComb(2004)]{mccomb} McComb, W. D., 2004, {\it Renormalization
%%%Group, A Guide for Beginners}, Clarendon Press, Oxford
%%
\item[]{} Meakin, C., 2006, Ph. D. Dissertation, Steward
Observatory, University of Arizona, Tucson AZ
%%
%\item[]{} Meakin, C., \& Arnett, D.,
%\apj, 637, 53 %(C and O shell)
%%
%\item[]{} Meakin, C., \& Arnett, D.,
%2007a, \apj, 665, 690 %(anelastic)
%%
\item[]{} Meakin, C., \& Arnett, D.,
2007b, \apj, 667, 448 %(I. hydro)
\item[]{} Meakin, C., \& Arnett, D., 2010,
\apss, 328, 221
\item[]{} Meakin, C., \& Arnett, D., 2011, \apj, 733, 78
%%
%%\bibitem[Meakin et al.(2011)]{mag11} Meakin, C., Arnett, D., \& Gore, ?,
%%2011, in preparation

\item[]{} Michaud, G., Richard, O.,
Richer, J., \& VandenBerg, D. A. 2004, \apj, 606, 452
\item[]{} Miesch, M. S., 2005, Living Reviews in Solar Physics, 2, 1
%%
%%\bibitem[Mo\'cak, \etal(2010)]{mocak10} Mo\'cak, M., Campbell, S. W.,
%%M\"uller, E., Kifonidis, K., 2010, \aap, 510, 114
%%%
%%\bibitem[Mo\'cak, \etal(2009)]{mocak09} Mo\'cak, M., M\"uller, E., Weiss, A.,
%%Kifonidis, K., 2009, \aap, 501, 659
%%%
%%\bibitem[Mo\'cak, \etal(2008)]{mocak08} Mo\'cak, M., M\"uller, E., Weiss, A.,
%%Kifonidis, K., 2008, \aap, 490, 265
\item[]{} Mo\'cak, M., Siess, L., \& M\"uller, E., 2011, \aanda, 533, A53
%%%
\item[]{} Moc{\'a}k, M., Meakin, C., Viallet, M., \& Arnett, D.\ 2014, arXiv:1401.5176
%%\bibitem[Monin \& Yaglom(1971)]{monin} Monin, A. S. \& Yaglom, A. M.,
%%1971, {\it Statistical Fluid Mechanics: Mechanics of Turbulence}, vol. 1,
%%Dover Publications, Mineola NY
%%
%%%\bibitem[Morton(1996)]{m96} Morton, K. W., 1996, Numerical Solution
%%%of Convection-Diffusion Problems, Chapman \& Hall, London
%%
\item[]{} Murphy, J. W.,
Burrows, A., \& Heger, A., 2004, \apj, 615, 460 %(check date)

\item[]{} Murphy, J. W., \& Meakin, C., 2011, \apj, 742, 74 

\item[]{} Pope, S. B., 2000, {\it Turbulent Flows}, 
Cambridge University Press, Cambridge, GB
%%
%%%\bibitem[Popper(1987)]{pop87}Popper, D. M. 1987, \apj, 313, L81
%%
%%\bibitem[Porter \etal(2002)]{porter02} Porter, D. H., Pouquet, A.,
%%\& Woodward, P. R., \pre, 66, 026301
%%
%%\bibitem[Porter \etal(1999)]{porter99} Porter, D. H., Pouquet, A.,
%%Sytine, I.,\& Woodward, P. R., Physica A, 263, 263
%%
%%\bibitem[Porter et al.(2001)]{pw01}Porter, D., Woodward, P., Toomre,
%%J., \& Brummel, N. H. 2001, www.lcse.umn.edu/MOVIES
%%
%%\bibitem[Porter \& Woodward(1994)]{pw94} Porter, D. H., \& Woodward, P. R.,
%%1994, \apjs, 93, 309
%%
%%\bibitem[Porter \& Woodward(2000)]{pw00} Porter, D. H., \& Woodward, P. R.,
%%2000, \apjs, 127, 159
%%
%%\bibitem[Porter, Woodward, \& Jacobs(2000)]{pwj00} Porter, D. H.,  
%%Woodward, P. R., \& Jacobs, M. L., 2000, Ann. N. Y. Acad. Sci., 898, 1 
%%(Proceedings of 
%%Fourteenth International Annual Florida Workshop in Nonlinear Astronomy 
%%and Physics, {\it Astrophysical Turbulence and Convection}, University 
%%of Florida, Feb. 1999).
%%
%%%\bibitem[Press et al.,(1992)]{pre92} Press, W. H., Teukolsky, S. A.,
%%%  Vetterling, W. T., \& Flannery, B. P., 1992, Numerical Recipes in
%%%  FORTRAN, Second Edition, University Press: Cambridge
%%
%%\bibitem[Press(1981)]{press81} Press, W. H. 1981, \apj, 245, 286
%%
%%\bibitem[Press \& Rybicki(1981)]{pr81} Press, W. H. \& Rybicki,
%%G. 1981, \apj, 248, 751

\item[]{} Quataert, E., \& Shiode, J., 2012, \mnras, 423, L92
\item[]{} Rakavy, G., Shaviv, G.,\& Zinamon, Z., 1967, \apj, 150, 131
%%
%%%\bibitem[Randich et al.(2001)]{ran01} Randich, S., Pallavicini, R.,
%%%Meola, G., Stauffer, J. R., \& Balachandran, S. C. 2001, \aap, 372,
%%%862
%%
%%%\bibitem[Rauscher \& Thielemann(2000)]{rt00}Rauscher, T., \& 
%%%  Thielemann, K.-F., 2000, Atomic Data Nuclear Data Tables, 75, 1
%%
%%%\bibitem[Reif(1965)]{r65} Reif, F. 1965, {\it Fundamentals of
%%%Statistical and Thermal Physics}, McGraw-Hill Book Co., N.Y.
%%
%%%\bibitem[Ribas et al.(2000)]{rib00}Ribas, I., Jordi, C., Torra, J., \&
%%%Gim\'{e}nez, \'{A}. 2000, \mnras, 313, 99
%%
%%%\bibitem[Robinson et al.(2004)]{rob04} Robinson, F. J., Demarque, P.,
%%%Li, L. H., Sofia, S., Kim, Y.-C., Chan, K. L., \& Guenther,
%%%D. B. 2004, \mnras, 347, 1208
%%
%%%\bibitem[Rogers \& Nayfonov(2002)]{rn02} Rogers, F. J., \& Nayfonov, A., 2002,
%%%\apj, 576, 1064
%%
%%%\bibitem[Rogers \& Glatzmaier(2005a)]{rg05a} Rogers, T. M., \& Glatzmaier, 
%%%G. A., \mnras, 364, 1135
%%
%%%\bibitem[Rogers \& Glatzmaier(2005b)]{rg05b} Rogers, T. M., \& Glatzmaier, 
%%%G. A., \apj, 620, 432
%%
%%%\bibitem[Ruelle \& Takens(1971)]{rt71} Ruelle, D., \& Takens, F., 1971,
%%%Commun. Math. Phys. 20, 167
%%
%%%\bibitem[Samadi, \etal(2006)]{sam06} Samadi, Kupka, F., R., Goupil, M. J.,
%%%Lebreton, Y., \& van't Veer-Menneret, C., 2006, \aap, 445, 233
%%
%%%\bibitem[Saslaw \& Schwarzschild(1965)]{ss65} Saslaw, W. C. \& 
%%%Schwarzschild, M., 1965, \apj, 142, 1468
%%
%%%\bibitem[Schatzman(1999)]{every}Schatzman, E., 1999, \apss, 265, 97
%%
%%%\bibitem[Schmelz, et al.(2005)]{schmelz} Schmelz, J. T., Nasraoui, K., 
%%%Roames, J. K., Lippner, L. A., \& Garst, J. W., 2005, \apjl, 634, L197
%%
\item[]{} Schwarzschild, M., 1975, \apj, 195, 137

%%%\bibitem[Seaton(2005)]{ms05}Seaton, Mike, as quoted by Bahcall, J. N., 
%%%Physics World, 2005, February, 266
%%
%%%\bibitem[Sedov(1997)]{sedov} Sedov, L. I., {\it Mechanics of Continuous Media},
%%%World Scientific, Singapore
%%
%%%\bibitem[Shannon(1948)]{shannon} Shannon, C., 1948, Bell System 
%%%Technical Journal, 27, 379
%%
%%%\bibitem[Shaviv \& Salpeter(1973)]{ss73} Shaviv, G. \& Salpeter, E. E.,
%%%1973, \apj, 184, 191

\item[]{} Shiode, J. H. \& Quataert, E., 2013, arXiv:1308.5878v1, \apj, submitted

\item[]{} Smith, N., \& Arnett, W. D., 2013, arXiv1307.5035S, \apj, submitted

%%%\bibitem[Soderblom et al.(1993)]{sod93} Soderblom, D. R., Jones,
%%%B. F., Balachandran, S., Stauffer, J. R., Duncan, D. K., Fedele,
%%%S. B., \& Hudon, J. D. 1993, \apj, 106, 1059
%%
%%\bibitem[Spiegel(1971)]{spie71} Spiegel, E. 1971, \araa, 9, 323
%%
%%%\bibitem[Spiegel(1972)]{spie72} Spiegel, E. 1972, \araa, 10, 261
%%
\item[]{} Spitzer, L., 1962, {\it Physics of
Fully Ionized Gases}, Second Edition, Interscience Publishers, NY
%%
%%%\bibitem[Smagorinsky(1963)]{smag63} Smagorinsky, J., 1963, Mon. 
%%%Weather Rev., 91(3), 99
%%
%%%\bibitem[Smith, J. D., \etal(2009)]{jds09} Smith, J. D. T., Rudnick, L., Delaney, T.,
%%%Rho, J, Gomez, H., Kozasa, T., Reach, W., Isense, K., 2009, \apj, 693, 713
%%
%%%\bibitem[Smith, N.,(2010)]{nathan10} Smith, Nathan, 2010, in {\it Active OB Stars:
%%%structure, evolution, mass-losss, and critical limits}, Proc. IAU Symposium 272,
%%%ed. C. Neiner, G. Wade, G. Meynet, \& G. Peters

\item []{} Smith, N., \& Arnett, W. D., 2013, arXiv:1307.5035S, submitted to \apj

%%
%%%\bibitem[Smith \& Woodruff(1998)]{sw98} Smith, L. M., \& Woodruff, S. L.,
%%%1998, Ann. Rev. Fluid Mech., 30, 275

\item[]{} Stancliffe, R. J., Deearborn, D. S. P., Latttanzio, F. C., Heeap, S. A., \& Campbell, S. W., 2011, \apj, 742, 121
%%
%%%\bibitem[Stassun et al.(2004)]{sta04} Stassun, K. G. Stassun, Mathieu,
%%%R. D., Vaz, L. P. R. , Stroud, N., \& Vrba, F. J. 2004, \apjs, 151,
%%%357
%%
%%%\bibitem[Stauffer et al.(1998)]{ssk98} Stauffer, J. R., Schultz, G.,
%%%\& Kirkpatrick, J. D. 1998, \apj, 499, L199
%%
%%%\bibitem[Stauffer et al.(1999)]{aper99} Stauffer, J. R., Barrado y
%%%Navascu\'{e}s, D., Bouvier, J., Morrison, H. L., Harding, P., Luhman,
%%%K. L., Stanke, T., McCaughrean, M., Terndrup, D. M., Allen, L., \&
%%%Assouad, P. 1999, \apj, 527, 219
%%
\item[]{} Stellingwerf, 1984, \apj, 284, 712
%%%\bibitem[Stein(1967)]{s67} Stein, R. F. 1967, \solphys, 2, 385
%%
%\bibitem[Stein \& Nordlund(1998)]{sn98} Stein, R. F., \& Nordlund, A., 1998,
%\apj, 499, 914
%%
%%\bibitem[Stein, et al.(2008)]{sngbs08} Stein, R. F., Nordlund, \AA, Georgobiani, D., Benson, D.,
%%Schaffenberger, W., in proceedings GONG2008/SOHO21, arXiv0811.0472
%%
%%%\bibitem[Stickland, Koch \& Pfeiffer(1992)]{skp92} Stickland, D. J.,
%%%  Koch, R. H., \& Pfeiffer, R. J., 1992, Obs., 112, 277
%%
%%%\bibitem[Stickland, Lloyd, \& Corcoran(1994)]{slc94} Stickland, D. J.,
%%%  Lloyd, C., Corcoran, M. F., 1994, Obs., 114, 284
%%
%%\bibitem[Stone \& Gardiner(2010)]{sg10} Stone, J. M., \& Gardiner, T. A., 2010,
%%\apjs, 189, 142
%%
%%\bibitem[Sytine, \etal(2000)]{sytine} Sytine, I., Porter, D., Woodward, P.,
%%Hodson, S. W., \& Winkler, K-H., 2000, \jcp, 158, 225
%%
%%%\bibitem[Talon, Kumar, \& Zahn(2002)]{tkz02} Talon, S., Kumar, P.,
%%%\& Zahn, J-P., 2002, \apj, 574, L175
%%
%%%\bibitem[Tassoul(1978)]{tas} Tassoul, J.-L. 1978, {\it Theory of
%%%Rotating Stars}, Princeton University Press, Princeton NJ
%%
%%\bibitem[Tassoul(2000)]{tas} Tassoul, J.-L. 2000, {\it Stellar
%%Rotation}, Cambridge University Press, NY
%%
\item[]{} Tennekes, H., \& Lumley, 
J. L., 1972, {\it A First Course in Turbulence}, MIT Press, Cambridge MA
%%
%%
%%%\bibitem[Thompson \& Steward(1986)]{thomp} Thompson, J. M. T. \& 
%%%Stewart, H. B., 1986, {\it Nonlinear Dynamics and Chaos}, John Wiley
%%%and Sons, New York
%%
%%%\bibitem[Thorburn et al.(1993)]{thor93} Thorburn, J. A., Hobbs,
%%%L. M., Deliyannis, C. P., \& Pinsonneault, M. H. 1993, \apj, 415, 150
%%
\item[]{} Thoul, A. A., Bahcall, J. N., \& Loeb, A. 1994, \apj, 421, 828
%%
%%%\bibitem[Timmes \& Arnett(1999)]{ta99} Timmes, F. X. \& Arnett, D., 1999,
%%%\apjs, 125, 277
%%%
%%%\bibitem[Timmes \& Swesty(2000)]{ts00} Timmes, F. X. \& Swesty, F. D.
%%%2000, \apjs, 126, 501
%%
%%%\bibitem[Tritton(1988)]{tritton} Tritton, D. J., {\it Physical Fluid
%%%Dynamics}, 2nd ed., Oxford University Press, Oxford UK
%%
%%\bibitem[Turner(1973)]{turner73} Turner, J. S., 1973, {\it Buoyancy Effects
%%in Fluids}, Cambridge University Press, Cambridge UK
%%
\item[]{} Unno, W., Osaki, Y., Ando, H., Saio, H., \&
Shibahashi, H., 1989, {\it Nonradial Oscillations of Stars}, 2nd. ed.,
University of Tokyo Press, Tokyo
%%
%%%\bibitem[van den Heuvel(2010)]{vdh10} van den Heuvel, E. P. J., 2010,
%%%{\it New Astronomy Reviews}, 54, 140
%%
\item[]{} Viallet, M., Meakin, C., Arnett, D., Mocak, M.
2013, \apj, 769, 1
\item[]{} Vitense, E., 1953, \zap, 32, 135
\item[]{} von Neumann, J., 1948, in \textit{Collected Works,
Volume~VI}, 1963, Pergamon Press, Oxford, p. 467-9
%%
%%%\bibitem[Wallerstein(1973)]{w73} Wallerstein, G. 1973, \araa, 11, 115
%%
%%%\bibitem[Wallerstein \& Knapp(1998)]{wk98} Wallerstein, G, \& Knapp,
%%%G. R. 1998, \araa, 36, 369
%%
\item[]{} Weaver, T., Zimmerman, G.,
\& Woosley, S., 1978, \apj, 225, 1021
%%
%%%\bibitem[Wegner et al.(1991)]{weg91} Wegner, G., Reid, I. N., \&
%%%McMahan, R. K. Jr. 1991, \apj, 376, 186
%%
%%%\bibitem[Wilson(1970)]{kwilson} Wilson, K. G., 1970, \prd 2, 1438
%%
%%%\bibitem[Wongwathanarat, Janka, \& M\"uller(2010)]{wjm10} Wongwathanarat, A.,
%%%Janka, H.-T., \& M\"uller, E., 2010, \apj, 725, 106
%%
%%\bibitem[Woodward, Porter, \& Jacobs(2003)]{wpj03} Woodward, P. R.,
%%Porter, D. H., \& Jacobs, M., 2000, in {\it 3D Stellar Evolution},
%%ed. S. Turcotte, S. C. Keller, \& R. M. Cavallo, 
%%ASP Conference Series 293, p. 45
%%
%%\bibitem[Woodward et al.(2006)]{wood06} Woodward, P. R., Porter, D., Anderson, S.,
%%\&  Fuchs, T., 2006, in {\it Numerical Modeling of Space Plasma Flows}, 
%%ed. N. V. Pogorelov \& G. P. Zank, Astron. Soc. Pacific Conf. Series, 359, 97
%%
%%\bibitem[Woodward(2007)]{woodward} Woodward, P. R., 2007, in {\it Implicit
%%Large Eddy Simulantions}, ed. F. F. Grinstein, L. G. Margolin, \& W. J.
%%Rider, Cambridge University Press, p. 130 
\item[]{} Woodward, P. R., Herwig, F., \& Lin, P-H., 2014, \apj (submitted), arXiv:1307.3821v1
%%
%%\bibitem[Woosley, Heger, \& Weaver(2002)]{whw02} Woosley, S. E., Heger, A., 
%%\& Weaver, T. A., 2002, Rev. Mod. Phys., 74, 1015
%%
\item[]{} Woosley, S. E., \& Heger, A., 2007, Phys. Rep., 442, 269

%%%\bibitem[Yakhot \& Orszag(1986)]{yakhot} Yakhot, V., \& Orszag, S. A., 1986,
%%%\prl 57, 1722
%%
%%%\bibitem[Young \& Arnett(2005)]{ya05} Young, P. A.
%%%\& Arnett, D. 2005, \apj, 618, 908
%%
%%\bibitem[Young et al.(2003)]{ykra} Young, P. A., Knierman, K. A.,
%%Rigby, J. R., \& Arnett, D. 2003, \apj, 595, 1114
%%
%%%\bibitem[Young, Mamajek, Arnett, \& Liebert(2001)]{ymal} Young, P. A., 
%%%Mamajek, E.~E., Arnett, D., \& Liebert, J. 2001, \apj, 556, 230
%%
%%%\bibitem[Young(2005)]{pry05} Young, P. R., 2005,
%%%\aap, 444, L45
%%
%%%\bibitem[Young, etal(2005)]{yetal05} Young, P. A., Meakin, C.,
%%%Arnett, D., Fryer, C., 2006, \apj, 629, L101
%%
%%%\bibitem[Young, etal(2006)]{yetal06} Young, P. A., Fryer, C., Hungerford, A.,
%%%Arnett, D., Rockefeller, G., Timmes, F. X., Voit, B., Meakin, C.,
%%%Eriksen, K. A., 2006, \apj, 640, 891
%%
%%%\bibitem[Young, \etal(2008)]{yeafr08} Young, P. A., Ellinger, C., Arnett, D., Fryer, C.,
%%%Rockefeller, G., 2008, in Proc. 10th Symposium on Nuclei in the Cosmos.
%%
%%\bibitem[Zahn(1992)]{zahn92} Zahn, J.-P., 1992, \aap, 265, 115
%%
%%\bibitem[Zeldovich, et al.(1985)]{zeld85} Zeldovich, Ya. B., Barenblatt, G. I.,
%%Librovich, V. B., \& Makhviladze, G. M., The Mathematical Theory of Combustion
%%and Explosions, Plenum, New York.

\end{itemize}

%\end{thebibliography}
%

\end{document}